\documentclass{emulateapj}
\usepackage{longtable}
\usepackage{graphicx}
\usepackage{graphics}        
\usepackage[english]{babel}
\usepackage{url}
\usepackage{amssymb}
\usepackage{amsmath}
\usepackage{wasysym}
\usepackage{color}
\usepackage{rotating}
\usepackage{array}
\usepackage{float}
\usepackage{natbib}

\shorttitle{High-redshift galaxy clusters around radio-loud AGN}
\shortauthors{Wylezalek et al.}

\begin{document}

\title{Galaxy Clusters around radio-loud AGN at $1.3<z<3.2$ as seen by \textit{Spitzer}}

\author{Dominika Wylezalek\altaffilmark{1,2}, Audrey Galametz\altaffilmark{3}, Daniel Stern\altaffilmark{1}, Jo\"{e}l Vernet\altaffilmark{2}, Carlos De Breuck\altaffilmark{2}, Nick Seymour\altaffilmark{4}, Mark Brodwin\altaffilmark{5}, Peter R. M. Eisenhardt\altaffilmark{1}, Anthony H. Gonzalez\altaffilmark{6},  Nina Hatch\altaffilmark{7}, Matt Jarvis\altaffilmark{8,9,10}, Alessandro Rettura\altaffilmark{11}, Spencer A. Stanford\altaffilmark{12, 13}, Jason A. Stevens\altaffilmark{9}}

\altaffiltext{1}{Jet Propulsion Laboratory, California Institute of Technology, 4800 Oak Grove Dr., Pasadena, CA 91109, USA}
\altaffiltext{2}{European Southern Observatory, Karl-Schwarzschildstr.2, D-85748 Garching bei M\"{u}nchen, Germany}
\altaffiltext{3}{INAF - Osservatorio di Roma, Via Frascati 33, I-00040, Monteporzio, Italy}
\altaffiltext{4}{CASS, PO Box 76, Epping, NSW, 1710, Australia}
\altaffiltext{5}{Department of Physics and Astronomy, University of Missouri, 5110 Rockhill Road, Kansas City, MO 64110, USA}
\altaffiltext{6}{Department of Astronomy, University of Florida, Gainesville, FL 32611, USA}
\altaffiltext{7}{School of Physics and Astronomy, University of Nottingham, University Park, Nottingham, NG7 2RD, UK}
\altaffiltext{8}{Astrophysics, Department of Physics, Keble Road, Oxford OX1 3RH, UK}
\altaffiltext{9}{Centre for Astrophysics Research, STRI, University of Hertfordshire, Hatfield, AL10 9AB, UK}
\altaffiltext{10}{Physics Department, University of the Western Cape, Bellville 7535, South Africa}
\altaffiltext{11}{Cahill Center for Astrophysics, California Institute of Technology, Pasadena, CA 91125, USA}
\altaffiltext{12}{Physics Department, University of California, Davis, CA 95616, USA}
\altaffiltext{13}{Institute of Geophysics and Planetary Physics, Lawrence Livermore National Laboratory, Livermore, CA 94550, USA}

\begin{abstract}
We report the first results from the Clusters Around Radio-Loud AGN (CARLA) program, a Cycle 7 and 8 {\it Spitzer Space Telescope} snapshot program to investigate the environments of a large sample of obscured and unobscured luminous radio-loud AGN at $1.2 < z < 3.2$.  These data, obtained for 387 fields, reach
3.6 and 4.5 $\mu$m depths of [3.6]$_{\rm AB} = 22.6$ and [4.5]$_{\rm AB} = 22.9$ at the 95\% completeness level, which is two to three times fainter than $L^*$ in this redshift range. By using the color cut [3.6]-[4.5] $> -0.1$ (AB), which efficiently selects high-redshift ($z > 1.3$) galaxies of all types, we identify galaxy cluster member candidates in the fields of the radio-loud AGN. The local density of these IRAC-selected sources is compared to the density of similarly selected sources in blank fields. We find that 92\% of the radio-loud AGN reside in environments richer than average. The majority (55\%) of the radio-loud AGN fields are found to be overdense at a $\geqslant$ 2 $\sigma$ level; 10\% are overdense at a $\geqslant$ 5 $\sigma$ level. A clear rise in surface density of IRAC-selected sources towards the position of the radio-loud AGN strongly supports an association of the majority of the IRAC-selected sources with the radio-loud AGN. Our results provide solid statistical evidence that radio-loud AGN are likely beacons for finding high-redshift galaxy (proto-)clusters. We investigate how environment depends on AGN type (unobscured radio-loud quasars vs. obscured radio galaxies), radio luminosity and redshift, finding no correlation with either AGN type or radio luminosity. We find a decrease in density with redshift, consistent with galaxy evolution for this uniform, flux-limited survey.  These results are consistent with expectations from the orientation-driven AGN unification model, at least for the high radio luminosity regimes considered in this sample.
\end{abstract}


\section{Introduction}

The Infrared Array Camera \citep[IRAC;][]{Fazio_2004} of the {\it
Spitzer Space Telescope} is an incredibly sensitive tool for finding
and studying massive galaxies at high redshift.  For stellar
populations formed at high redshift, negative $k$-corrections provide
a nearly constant $4.5 \mu$m flux density over a wide redshift range
--- e.g., an $L^*$ galaxy formed at $z_f = 3$ will have [4.5] $\sim
17$ at $0.7 \leqslant z \leqslant 2.5$, which is sufficiently bright
that it is robustly seen in even 90~sec integrations with {\it
Spitzer}.  Several teams have been exploiting this capability to
identify large samples of galaxy clusters at $z \geqslant 1$
\citep[e.g.,][]{Eisenhardt_2008, Wilson_2009, Papovich_2010,
Galametz_2010b, Muzzin_2013}.  The {\it Spitzer} mid-infrared selection
is capable of finding both the very rare, massive evolved clusters at $z \geqslant 1.75$
\citep[e.g.,][]{Stanford_2012}, as well as the more numerous groups or
forming clusters at similarly high redshifts
\citep[e.g.,][]{Zeimann_2012, Gobat_2011}.
Teams are now using mid-IR-selected clusters for a range of important
studies such as using the colors, sizes, and mid-IR cluster galaxy
luminosity function to probe the formation epoch of massive cluster
galaxies \citep[e.g.][]{Mancone_2010, Mancone_2012, Rettura_2011,
Lidman_2012, Mei_2012, Snyder_2012}, measuring the galaxy cluster
correlation function out  to $z \sim 1.5$ \citep{Brodwin_2007},
probing evolution in the $\sigma - T_x$ correlation \citep{Brodwin_2011},
and targeted cosmological studies of $z > 1$ Type~Ia SNe in dust-free
cluster environments \citep{Dawson_2009, Suzuki_2012}.

Much of the work to date on high-redshift galaxy clusters comes
from field surveys, which provide both a strength and a weakness.
Uniformly selected galaxy cluster samples have the power to probe
basic cosmological parameters by measuring the growth of structure.
However, field surveys --- both in the mid-IR
\citep[e.g.,][]{Gettings_2012} and at other wavelengths \citep[e.g.,
Sunyaev-Zeldovich and X-ray surveys;][]{Vanderlinde_2010,
Fassbender_2011, Reichardt_2012} --- find few clusters at the highest
redshifts, $z \geqslant 1.5$.  While measuring the cluster space density
is required to use these systems as cosmological probes, many key galaxy cluster studies do
not require such knowledge \citep[e.g.,][]{Krick_2008, Galametz_2009, Suzuki_2012, Martini_2013}.

In order to efficiently identify the richest environments at yet higher
redshifts, targeted searches for high-redshift galaxy clusters have
many advantages. Towards that goal, literature that stretches back
nearly 50 years shows that powerful radio-selected AGN preferentially
reside in luminous red sequence galaxies \citep[e.g.,][]{Matthews_1965,
Best_1998, Venemans_2002, Hickox_2009, Griffith_2010}.  Indeed,
targeted searches for high-redshift clusters and proto-clusters
around powerful high-redshift radio galaxies (HzRGs) have proven
very successful and have a rich literature
\citep[e.g.,][]{Pentericci_2000, Stern_2003, Kurk_2004, Venemans_2007,
Doherty_2010, Galametz_2009,Galametz_2010, Hatch_2009, Matsuda_2011,
Mayo_2012}.

In a pilot study by our team, \citet{Galametz_2012} used a counts-in-cell analysis to identify overdensities of IRAC-selected high-redshift galaxy candidates in the fields of 72 HzRGs from the {\it Spitzer} High-Redshift Radio Galaxy program \citep[SHzRG;][]{Seymour_2007, Breuck_2010}. The HzRGs in that study spanned a range of $1.2 < z < 3$.  Using relatively shallow, $\sim 120$~s IRAC data, \citet{Galametz_2012} showed that radio galaxies preferentially reside in medium to dense regions, with 73\%\ of the targeted fields denser than average.  Apart from six newly discovered cluster candidates, several known (proto-)clusters were recovered as overdense fields in that analysis, and we have recently spectroscopically confirmed one of the more promising candidate fields as a new high-redshift cluster at $z = 2.02$ (Galametz et al., in prep.).

In this paper, we describe the first results from our $\sim 400$~hr {\it Warm Spitzer} snapshot program called Clusters Around Radio-Loud AGN,
or CARLA\footnote{As of March 2013, CARLA targets are still being scheduled; this manuscript details all observations obtained through February 2013.},  a Cycle~7 and Cycle~8 program which takes advantage of {\it Spitzer}'s impressive sensitivity to identify massive galaxies at high redshift.  To date, CARLA has targeted nearly 400 radio-loud AGN, including 187 radio-loud quasars (RLQs) and 200 HzRGs.  Targets were uniformly selected in radio luminosity over the redshift range $1.3 < z < 3.2$.  Similar to the successful pilot
study by \citet{Galametz_2012}, CARLA allows, for the first time, a sensitive, systematic study of the environments of a large sample
of powerful radio-loud sources over a wide redshift range.  We isolate high-redshift galaxy candidates using IRAC colors and
quantify the environments of the targeted AGN as compared to similar-depth blank-field surveys.  CARLA also allows us to investigate
how environment depends on radio-loud AGN properties such as redshift, radio luminosity and AGN class (e.g., type-1, or unobscured RLQs vs. type-2, or obscured HzRGs). A companion paper, Wylezalek et al. (in prep.), investigates the
luminosity function of CARLA-selected candidate galaxy cluster
members.

The paper is organized as follows: \S~2 describes how the sample of HzRGs and RLQs was selected and \S~3 gives details on the observations, data reduction, catalog generation and the completeness limit of the survey. We describe the analysis of the environments of the CARLA targets with respect to galaxy type, redshift and radio luminosity in \S~5 and draw conclusions from the results in \S~6.  Section 7 summarizes our results. Throughout the paper we assume $H_{0}$ = 70 km s$^{-1}$ Mpc$^{-1}$, $\Omega_{\rm{m}}$ = 0.3, $\Omega_{\Lambda}$ = 0.7. All magnitudes and colors are expressed in the AB photometric system unless stated otherwise.

\section{The CARLA Sample}

\begin{figure}
\centering
\includegraphics[scale = 0.71, trim = 0cm 0cm 0cm 3cm]{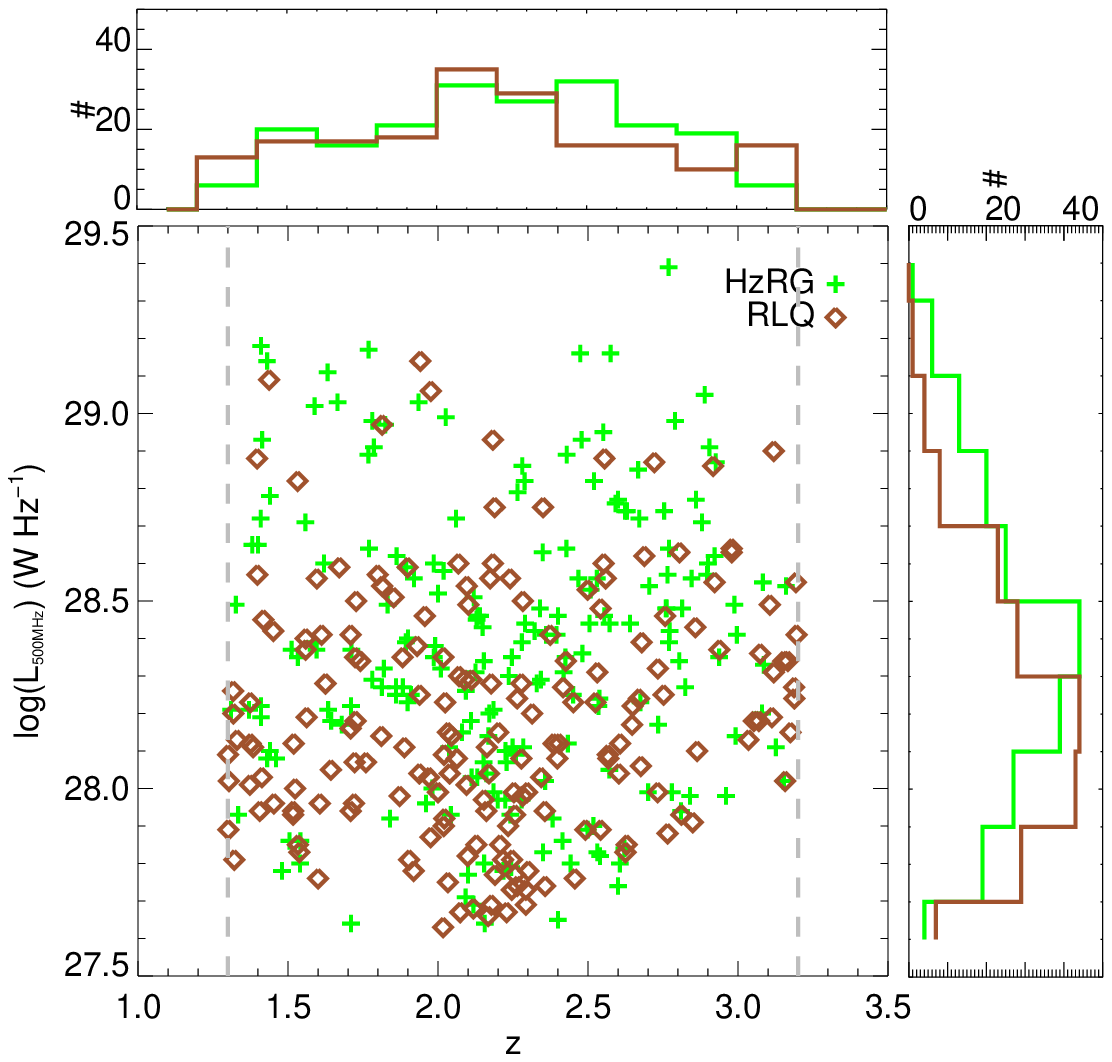}
\includegraphics[scale = 0.46, trim = 4cm 0cm 0.2cm 1cm]{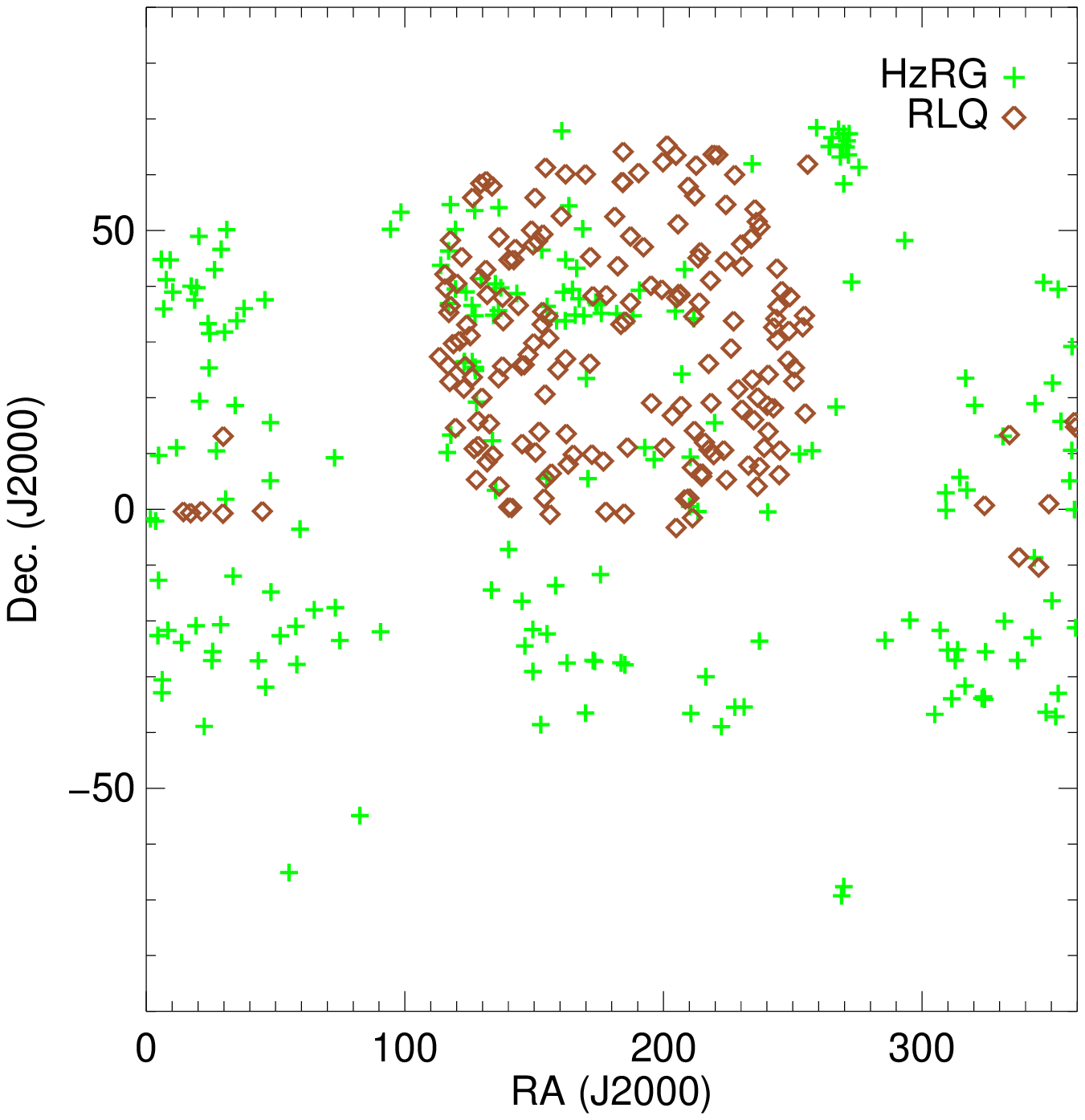}
\caption{\textit{Top panel:} Rest-frame 500 MHz luminosity vs. redshift. As shown by their distributions, both subsamples (HzRGs and RLQs) sample the targeted $L_{500\rm{MHz}} - z$ plane in a relatively uniform manner. \textit{Bottom panel:} Distribution of the CARLA radio-loud AGN across the sky. Most of the RLQs are chosen from the SDSS and are thus mainly located in the Northern hemisphere, while the HzRGs cover the full extragalactic sky. In both panels, green crosses show the HzRGs in our sample, while orange diamonds show the RLQs.}
\label{sample}
\end{figure}

A key goal of the CARLA survey is to investigate how environment
depends on AGN classification, and we therefore have targeted a
matched sample of unobscured (e.g., type-1) and obscured (e.g.,
type-2) radio-loud AGN.  Type-1 AGN, specifically broad-lined quasars
or RLQs for the high-luminosity, radio-loud sources targeted here,
show signatures of radiation originating from very close to the
central black hole.  In contrast, the central black hole is obscured
by dense absorbing material in type-2 AGN, or HzRGs for
the high-luminosity, high-redshift, radio-loud sample observed here.
In the standard AGN unification model \citep{Urry_1995} this
obscuration is produced by an optically thick circumnuclear torus
and differences in optical appearance only depend on the orientation
angle at which the AGN is observed, not due to intrinsic differences
between the two AGN types. A clear prediction is that both AGN types
should reside in identical environments.  Recently, an alternative
paradigm has been gaining attention in which type-2 AGN evolve into
type-1 AGN as their merger-triggered, dust-obscured AGN become more
powerful and clear out the environment \citep[e.g.][and references
therein]{Hopkins_2006}. This dynamical model would mean that type-1
and type-2 AGN trace different phases in the evolution of powerful
AGN, potentially associated with different environments.  On the
other hand, if the timescale associated with the AGN breaking out
of its obscuring cocoon is much shorter than the timescale on which
the environment develops, then no dependence of environment on AGN type
would be expected in the dynamical model.

The CARLA sample consists of 387 radio-loud AGN, including 187 RLQs
and 200 HzRGs (sample size as of 2013 February 20). The matched RLQ
and HzRG samples have been deliberately designed to be of similar
size with similar redshift and radio luminosity ranges. HzRGs are
defined as powerful, high-redshift radio sources with rest-frame
500 MHz radio luminosities $L_{500\rm{MHz}} \geqslant 10^{27.5}$ W
Hz$^{-1}$ and have optical spectra free of broad emission lines. 
This classification as narrow-line, i.e. type-2 AGN is based on the optical spectroscopy used to determine their redshifts. 
The HzRGs were selected from the updated compendium of
\citet*{Miley_2008} extended down to $z = 1.3$ using both flux-limited
radio surveys (e.g. MRC, 3C, 6C, 7C) and ultra-steep spectrum surveys
\citep[e.g.][]{Roettgering_1997, Breuck_2001}. The sample contains all
146 powerful HzRGs known as of 2010 at $2 < z < 3.2$.

To calculate $L_{500\rm{MHz}}$, we used the procedure described by \citet*{Miley_2008}. In short, we cross-correlated the HzRGs with the 1.4\,GHz NRAO VLA Sky Survey \citep[NVSS; ][]{Condon_1998} and the 74\,MHz VLA Low-Frequency Sky Survey \citep[VLSS; ][]{Cohen_2007}, using a 30\arcsec\ search radius. These two radio surveys cover a very large part of the extragalactic sky, and have well-matched spatial resolution, thereby avoiding concerns with missing flux. The rest-frame 500\,MHz frequency was deliberately chosen to ensure that the fluxes were interpolated rather than extrapolated between the two radio surveys. If no VLSS detection was available, we assumed a spectral index $\alpha = -1.1$ \citep[$S_{\nu} \propto \nu^{\alpha}$,][]{Miley_2008}.

To create the RLQ sample, we selected all optically bright ($M_{B} <
-26.5$) quasars in the Sloan Digitized Sky Survey \citep[SDSS;][]{Schneider_2010} and the 2dF QSO Redshift Survey \citep[2dFQZ; ][]{Croom_2004}, where the spectroscopy-based classifications were taken from the original catalogues. For the SDSS quasars the $M_{B}$ were taken from the original catalogue, and for the 2df quasars we calculated $M_{B}$ from the original catalogs assuming a quasar SED represented by a power law ($f_{\nu} \propto \nu^{\alpha_{RLQ}} , \alpha_{RLQ} = 0.5$ -- as used for the SDSS quasars) and we corrected for Galactic dust extinction.

To identify RLQs, we then correlated this sample with the NVSS using a 10\arcsec\ search
radius. As only $\sim$20\% of the RLQs are detected in the VLSS, we
could not calculate the radio spectral indices in a systematic
way. Moreover, RLQs may contain time-variable, Doppler beamed
emission, which would render the spectral indices from surveys
observed several years apart unreliable. We therefore assumed a
fixed $\alpha = -0.7$ for all RLQs to calculate their
$L_{500\rm{MHz}}$. We tested and justified this choice by computing
the mean and median value for the 133 out of 717 RLQs where the
VLSS-NVSS spectral indices could be derived, and find $-0.65$ and
$-0.74$, respectively. On average, the radio luminosities calculated
with a fixed $\alpha$ only differ from the precise radio luminosities
by $3$\% and this scatter will not have any influence on our results
and conclusions.

The final RLQ sample was obtained by considering only sources in
regions with low Galactic extinction ($E(B-V) \leqslant 0.1$) and by
matching to the redshift and radio power distribution of the HzRG
sample.  The upper panel in Fig. \ref{sample} shows rest-frame
$L_{500\rm{MHz}}$ vs.  redshift while the lower panel shows the
distribution of our targets across the sky. This sample selection
allows us to study evolutionary trends as well as trends with radio
luminosity for both types of AGN without any biases beyond the
selection being restricted to the most radio powerful sources.

\section{Data}

\subsection{Observations}

CARLA observed the lower redshift ($1.3 < z < 2$) sources with the \textit{Spitzer} IRAC camera at 3.6 and 4.5 $\mu$m (referred to as IRAC1 and IRAC2) during Cycle 7 with total exposure times of 800s in IRAC1 and 2000s in IRAC2. Slightly deeper observations were obtained for the higher redshift ($2 < z < 3.2$) portion of the sample during Cycle 8 with total exposure times of 1000s in IRAC1 and 2100s in IRAC2. IRAC employs 256 $\times$ 256 InSb detector arrays with 1\farcs22 pixels to map out a 5.2\arcmin\, $\times$ 5.2\arcmin\, field. The IRAC1 and IRAC2 data were obtained simultaneously over slightly offset fields, allowing CARLA to tailor the exposure times in each band for our intended science. Observations were all obtained with dithered 100~s observations.  In order to save the $\sim 5$~min overhead per target had the differing integration times for IRAC1 and IRAC2 been split across two Astronomical Observation Requests (AORs), we instead implemented this observing strategy in a single AOR using the the ``Fixed Cluster Offset'' mode in array coordinates. In AB magnitudes, the total exposure times give similar depths for both channels (see section 3.4) with the IRAC2 observations being slightly deeper. The total exposure times, which are slightly longer for the high-redshift portion of the sample (observed during Cycle 8), reach depths two magnitudes below $m^{*}$ over the whole redshift range (see Section 3.4). This allows us to robustly measure the faint-end slope of the cluster luminosity function out to high redshift, which is the focus of our second CARLA paper (Wylezalek et al., in prep.).

\subsection{Data Reduction}

The basic calibrated data were reduced and mosaicked using the MOPEX package \citep{Makovoz_2005} and resampled to a pixel scale of 0\farcs61. The MOPEX outlier (e.g., cosmic ray, bad pixel) rejection was optimized for the regions of deepest coverage in the center of the maps, corresponding to the location of the targeted AGNs. Due to artifacts and bright foreground stars, the maximum coverage is not reached in every part of the field. Regions with $\leqslant 85$\% of the maximum coverage were ignored for source extraction and further analysis.

\begin{figure*}
\centering
\includegraphics[scale = 0.45]{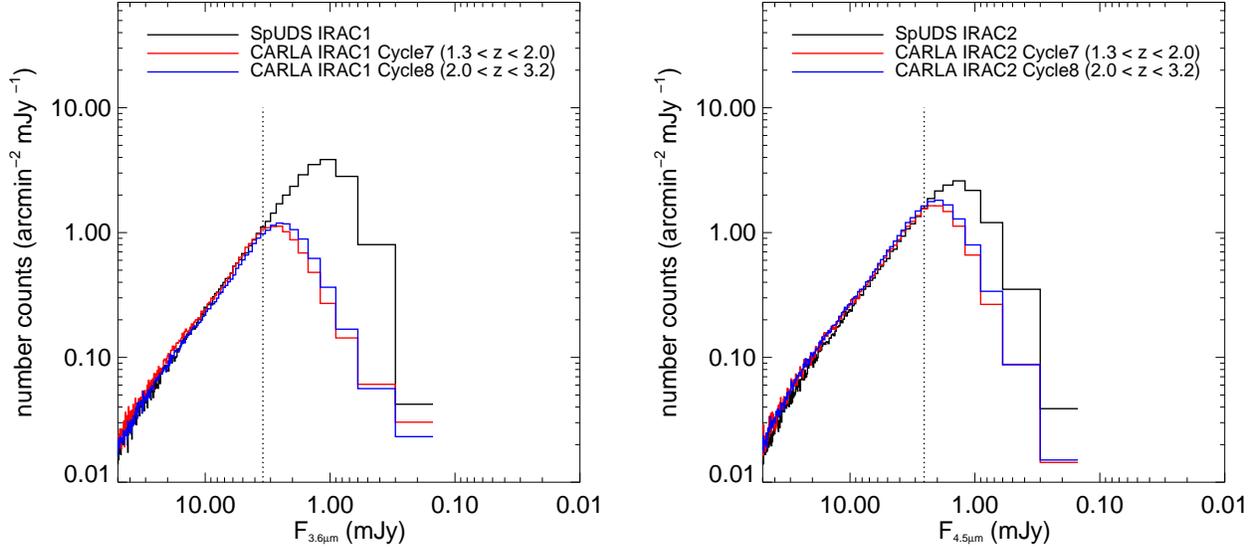}
\caption{Number counts for the SpUDS and CARLA surveys for sources detected at 3.6 $\mu$m (\textit{left}) and 4.5 $\mu$m (\textit{right}). These analyses rely on catalogs independently built for detecting sources in each band. The 95\% completeness limit in each band is derived by comparing the CARLA number counts to those from the SpUDS survey, which gives completeness limits of 3.45 $\mu$Jy for IRAC1 and 2.55 $\mu$Jy for IRAC2, as shown by the vertical dotted lines.}
\label{completeness}
\end{figure*}

\subsection{Source Extraction}

Source extraction was performed using SExtractor \citep{Bertin_1996} in dual image mode using the 4.5 $\mu$m frame as the detection image. We used the IRAC-optimized SExtractor parameters from \citet{Lacy_2005}. We measured flux densities in $4^{\prime\prime}$ diameter apertures, converting from the native MJy ster$^{-1}$ units of the images to $\mu$Jy pixel$^{-1}$ by multiplying with the conversion factor of 8.4615 $\mu$Jy pixel$^{-1}$ / (MJy ster$^{-1}$) for our 0\farcs61 pixel scale. We derived empirical aperture corrections from the curve of growth of bright, isolated stars in the CARLA survey. These multiplication factors, 1.42 for IRAC1 and 1.45 for IRAC2, correct the photometry to 24$^{\prime\prime}$-radius apertures. These values are systematically $\sim$ 17\% larger than the aperture corrections listed in the \textit{Spitzer} Instrument Handbook but are consistent with the aperture corrections found by \citet{Ashby_2009} for the \textit{Spitzer} Deep, Wide-Field Survey. The discrepancy with the Instrument Handbook is most likely due to differences in the background determination and slight differences in aperture radii \citep{Ashby_2009}.

\subsection{Completeness Limit}

We determined the 95\% completeness level by comparing the CARLA number counts to number counts from the \textit{Spitzer} UKIDSS Ultra Deep Survey (SpUDS, PI: J. Dunlop), a deep Cycle 4 \textit{Spitzer} Legacy program covering $~ 1\ \rm{deg}^{2}$ in the UKIDSS UDS field with IRAC and the Multiband Imaging Spectrometer aboard \textit{Spitzer} \citep[MIPS;][]{Rieke_2004}. SpUDS reaches greater sensitivities than CARLA, with IRAC1 and IRAC2 3 $\sigma$ depths of $\sim 1\ \mu$Jy (mag$_{\rm{AB}} \simeq 24$). Although SpUDS catalogs are publicly available, we performed source extraction on the final mosaics in the same way as for CARLA to ensure uniformity in our analyses. In particular, the public catalogs are cut at a depth of 8 $\sigma$, i.e. shallower than CARLA, and thus would not allow us to accurately measure the completeness of the CARLA catalogs. The aperture corrections we derive for bright, isolated sources in SpUDS are consistent with the ones derived for the CARLA survey. We created IRAC1 and IRAC2 SExtractor catalogs using single image mode. The differential number counts are shown in Fig. \ref{completeness}. As planned, the IRAC1 and IRAC2 observations match very well in depth. We determine a 95\% completeness limit of 3.45 and 2.55 $\mu$Jy relative to the SpUDS survey, corresponding to limiting magnitudes of [3.6] = 22.6 and [4.5] = 22.9 for the observations. These values are adopted as flux density cuts for all subsequent analyses.

\section{The Environments of Radio-Loud AGN}
\subsection{IRAC-Selected Sources}

\begin{figure*}
\centering
\includegraphics[scale = 0.4]{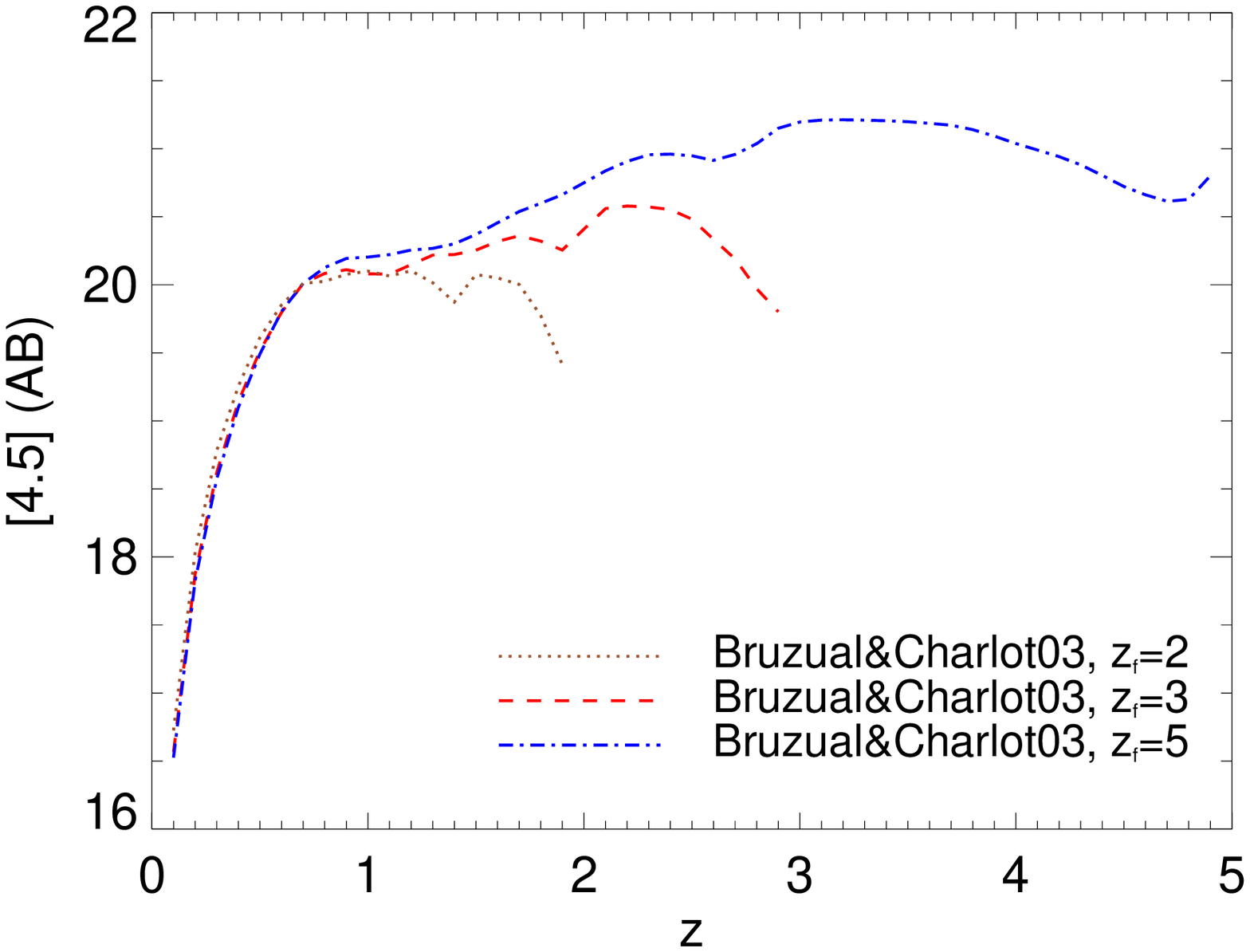}
\includegraphics[scale = 0.4]{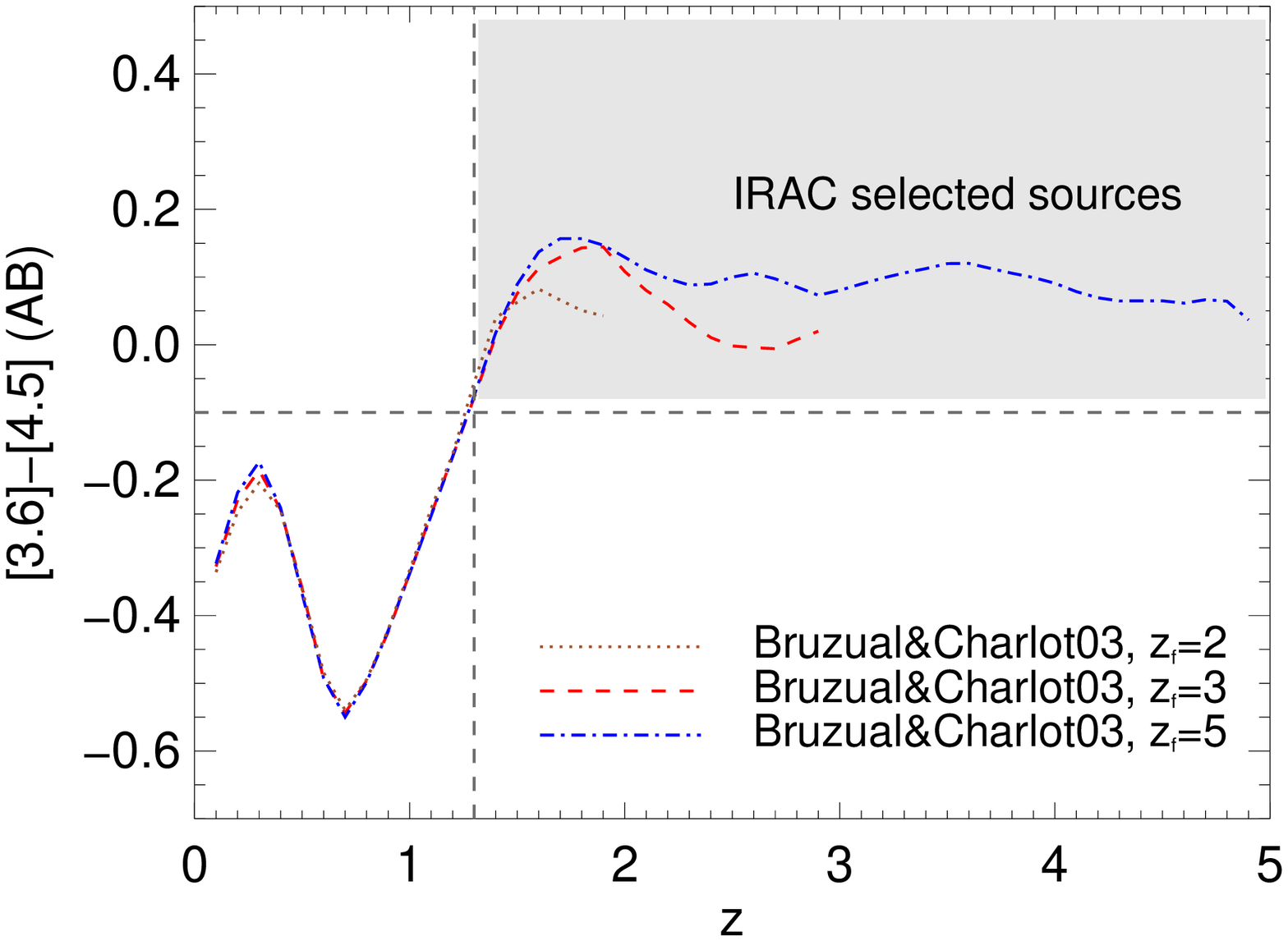}
\caption{Illustration of the IRAC color criterion. \textit{Left}: Predicted evolution of IRAC2 magnitude as a function of redshift. We plot the \citet{Bruzual_2003} models for different formation redshifts $z_{f}$ assuming a Salpeter initial mass function and a single exponentially decaying burst of star formation with $\tau = 0.1$ Gyr, normalized to the observed $m^{*}$ for galaxy clusters at $z = 0.7$, [4.5] = 16.75 \citep[Vega; ][]{Mancone_2010}. The models were generated with the model calculator EzGal \citep{Mancone_2012b}. A negative $k$-correction leads to an almost flat evolution of the IRAC2 magnitude at $z \geqslant 0.7$. \textit{Right}: The shift of the 1.6 $\mu$m bump across the IRAC bands leads to a red IRAC color for galaxies at $z > 1.3$.}
\label{irac_color}
\end{figure*}

We use a counts-in-cell analysis of color-selected IRAC sources to identify high-redshift overdense fields. \citet{Papovich_2008} showed that an IRAC color cut of [3.6]$-$[4.5] $> -0.1$ efficiently isolates galaxies regardless of age and galaxy type at $z > 1.3$. This criterion makes use of the 1.6 $\mu$m bump, a prominent spectral feature in the spectral energy distributions of galaxies that does not depend on the evolutionary stage of the galaxy. The bump is caused by a minimum in the opacity of the H$^{-}$ ion which is present in the atmospheres of cool stars \citep{John_1988} and is a commonly used indicator of redshift \citep[e.g., ][]{Simpson_1999, Sorba_2010}. The 1.6 $\mu$m bump enters the IRAC bands at $z \sim 1$ and provides a redder $[3.6]-[4.5]$ color for high-redshift galaxies as compared to low-redshift ($z<1$) galaxies (see Fig.\ref{irac_color}). Contamination can come from strongly star-forming galaxies at $0.2 < z < 0.5$ \citep{Papovich_2008}, cool brown dwarfs \citep{Stern_2007} and powerful AGN at all redshifts \citep{Stern_2005}.  However, these contaminations are not expected to be significant \citep[for a detailed discussion see ][]{Galametz_2012}. More than 90\% of objects satisfying this simple color criterion will be at $z > 1.3$ \citep{Papovich_2008}. In the following we refer to these high-redshift, mid-IR color-selected sources simply as `IRAC-selected sources'. 

\subsection{Comparison to Blank Fields}

\begin{figure}
\centering
\includegraphics[scale = 0.5]{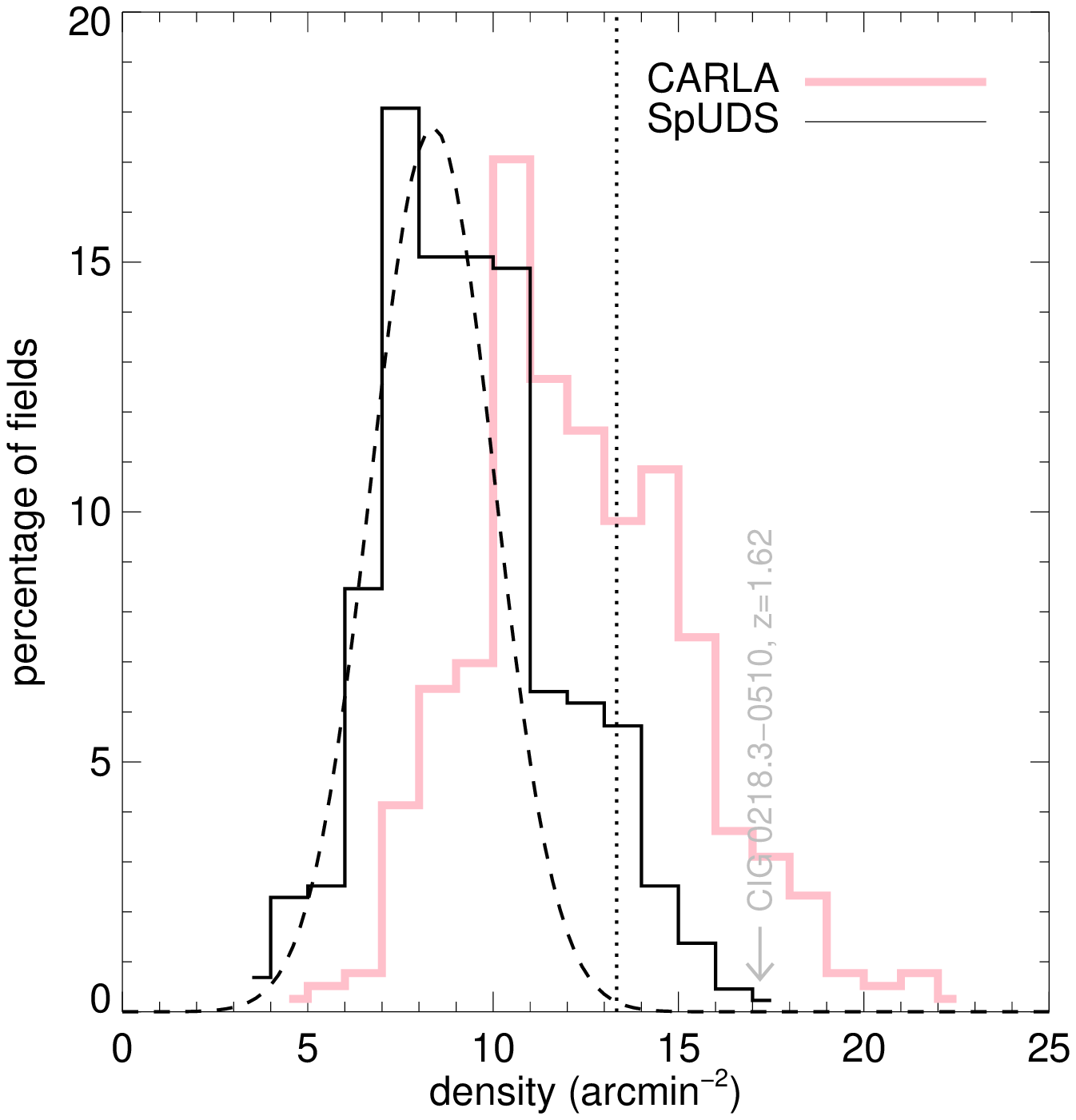}
\includegraphics[scale = 0.5]{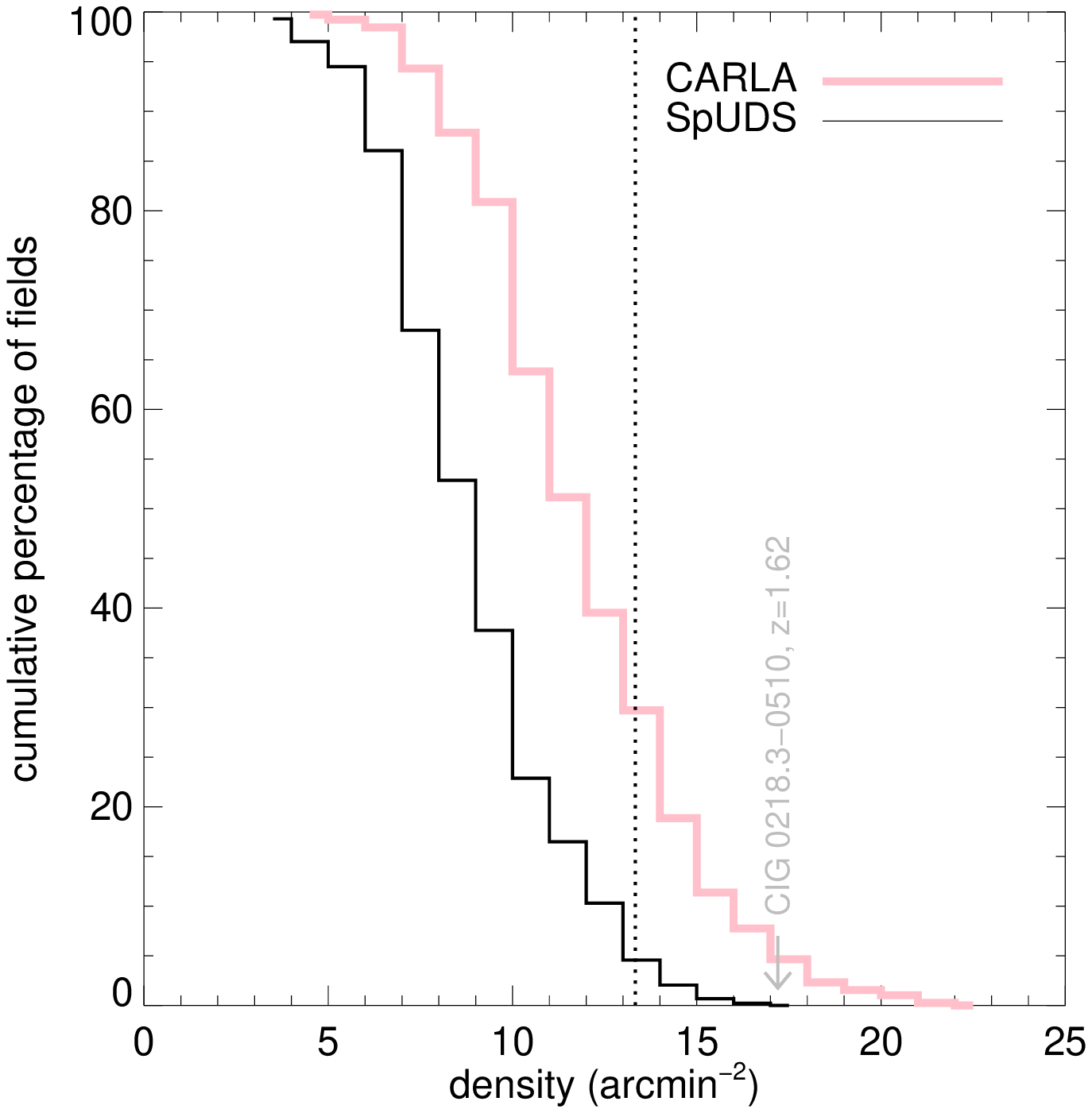}
\caption{\textit{Upper panel:} Histogram of the densities of IRAC-selected sources in the CARLA fields and the SpUDS survey. The Gaussian fit to the low-density half of the SpUDS density distribution is shown by the dashed black curve, giving $\Sigma_{\rm{SpUDS}} = 8.3\pm 1.6$ arcmin$^{-2}$. The vertical dotted line corresponds to $\Sigma_{\rm{SpUDS}} + 3\sigma$ density. We find that 37.0\% of the CARLA fields are overdense compared to this surface density while only 8.6\% of the SpUDS fields have surface densities of IRAC-selected sources above that value. \textit{Lower panel:} Cumulative density distribution of the CARLA and SpUDS fields. Only 18.7\% of the SpUDS fields are denser than $\Sigma_{\rm{SpUDS}}+2\sigma_{\rm{SpUDS}}$ in contrast to 55.3\% for the CARLA fields. Only 0.7\% of the SpUDS fields are denser than $\Sigma_{\rm{SpUDS}}+5\sigma_{\rm{SpUDS}}$ while 9.6\% of the CARLA fields are denser than this cut.}
\label{histo}
\end{figure}

\begin{figure}
\centering
\includegraphics[scale = 0.8]{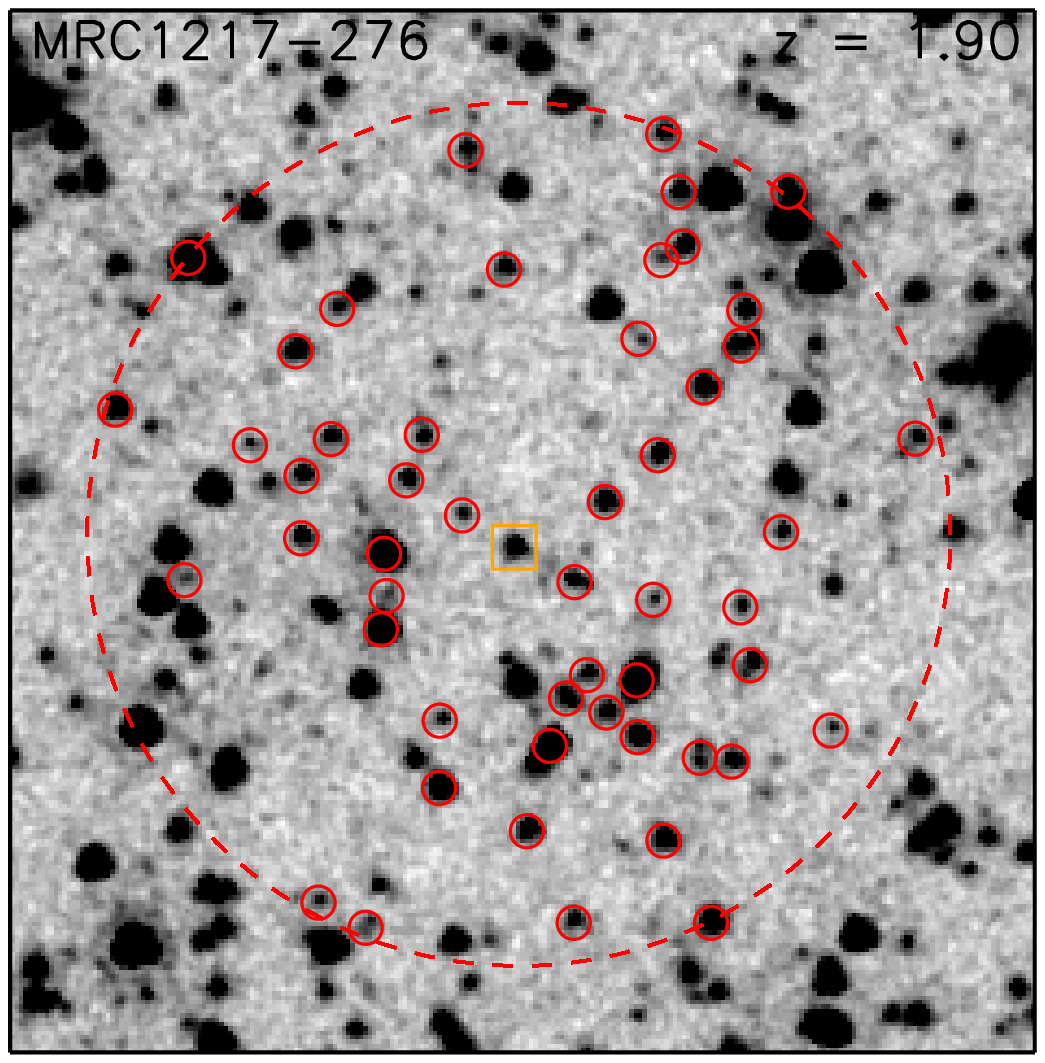}
\includegraphics[scale = 0.79]{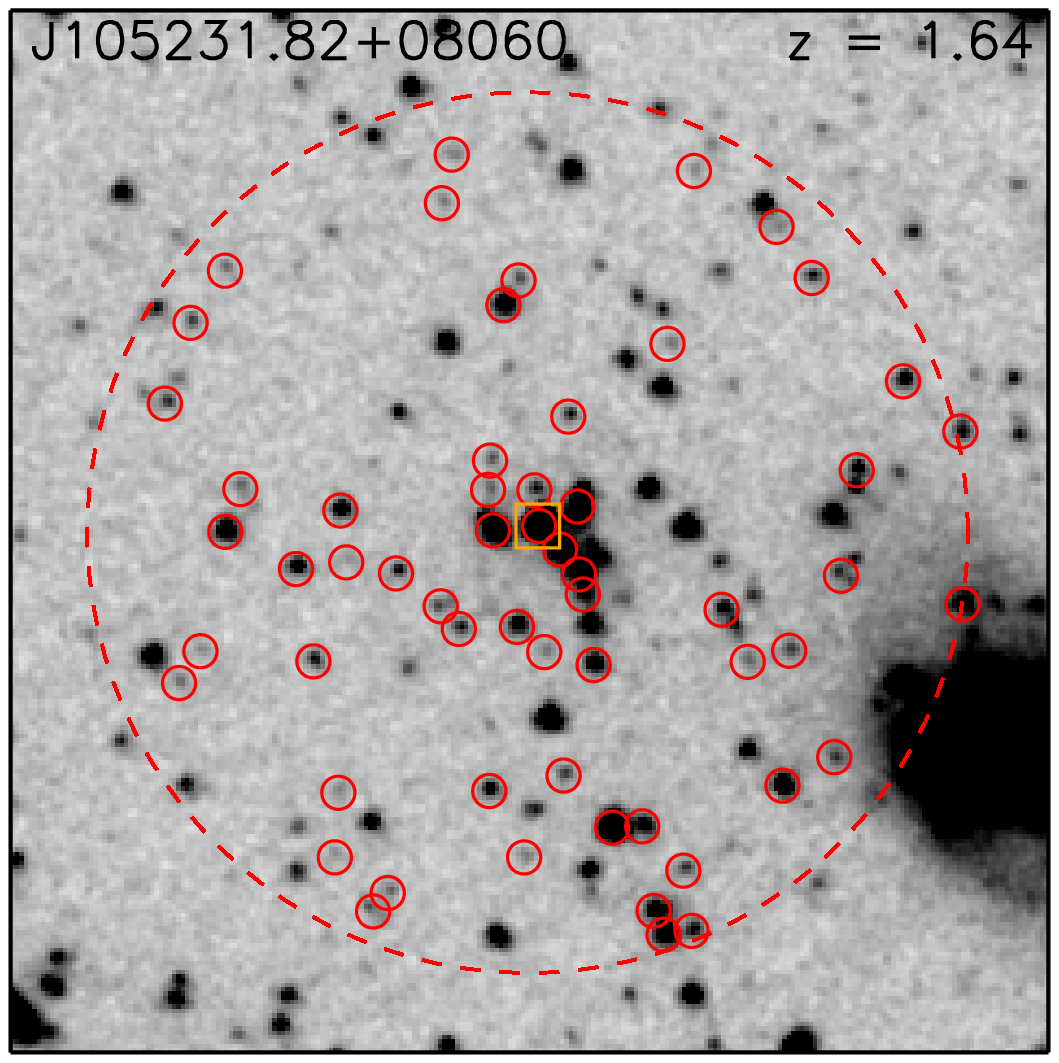}
\caption{Postage stamps of the fields of MRC1217-276 and J105231.82+08060, a radio galaxy and a radio-loud quasar at $z =1.90$ and $z = 1.64$, respectively. The IRAC color-selected sources within a 1 arcmin radius, shown with the large dashed circle, are marked with small solid red circles. The radio-loud AGN are marked by the orange squares. }
\label{data_img}
\end{figure}

\begin{table*}
\caption{The CARLA sample (in R.A. order).}
\label{tablecarla}
\begin{center}
\begin{tabular}{l c c c c c c}
\hline\hline
Name & R.A. & Dec. & $z$ & log($L_{500\rm{MHz}})$ & $\Sigma$& AGN Type \\
 & (J2000) & (J2000) & & (W Hz$^{-1}$) & (arcmin$^{-2}$)&  \\
&  & & & & & \\
\hline
USS0003-019	&	00:06:11.00	&	-01:41:50.2	&	1.54	&	27.86	&	10.5 & HzRG	\\	
PKS\_0011-023	&	00:14:25.00	&	-02:05:56.0	&	2.08	&	28.15	&	10.2 & HzRG	\\	
MRC\_0015-229	&	00:17:58.20	&	-22:38:03.8	&	2.01	&	28.32	&	15.6 & HzRG	\\	
BRL0016-129	&	00:18:51.40	&	-12:42:34.6	&	1.59	&	29.02	&	15.3 & HzRG	\\	
MG0018+0940	&	00:18:55.20	&	+09:40:06.9	&	1.59	&	28.39	&	13.4 & HzRG	\\
& & & \vdots & & & \\
\end{tabular}
\end{center}
%\newline
\centering{This table is published in its entirety in the electronic edition of ApJ; a portion is shown here for guidance regarding its form and content.}	
\end{table*}

We use the SpUDS survey to derive a mean blank field density of IRAC-selected sources. The catalogs have been cut at the CARLA depth, i.e. all sources fainter than the CARLA 95\% completeness limit are neglected. We then placed 436 non-overlapping circular cells of 1$^\prime$ radius onto the field and measured the density of IRAC-selected sources in these cells. None of the CARLA targets lie in the SpUDS survey.

We then measured the number of IRAC-selected sources in an aperture of 1$^\prime$ radius centered on the radio-loud AGN.  At $ 1< z < 3$, 1$^{\prime}$ corresponds to $\sim500$ kpc, matching typical cluster sizes for $z > 1.3$ mid-IR selected clusters with log$(M_{200}/M_{\astrosun}) \sim 14.2$ \citep[e.g. ][]{Brodwin_2011}. The surface density of IRAC-selected sources in the CARLA fields takes into account the low coverage areas (see Section 3.2). Table \ref{tablecarla} lists the names, positions, redshifts, radio luminosities and surface density of IRAC-selected sources in the CARLA fields sorted by right ascension. The full table is available online. The upper panel in Fig. \ref{histo} shows the distribution of surface densities for IRAC-selected sources in the SpuDS and CARLA fields. A perfect, structure-free blank field would show a Gaussian distribution of densities peaking at the mean blank field density. As expected, both the SpUDS and CARLA distributions are asymmetric with a high-density tail since even the $\sim 1$ deg$^{2}$ SpuDS field contains large scale structure. For example, a high-density tail of IRAC-selected sources is also seen in the Lockman Hole field \citep{Papovich_2008}. Following the methodology of \citet{Papovich_2008}, \citet{Galametz_2012} and \citet{Mayo_2012}, we fit a Gaussian to the low-density half of the distribution (i.e. the distribution of the lower density regions determined with respect to the maximum of the distribution) to determine the peak density $\Sigma_{\rm{SpUDS}} = 8.3$ arcmin$^{-2}$ and the standard deviation of the low-density half, $\sigma_{\rm{SpUDS}} = $ 1.6 arcmin$^{-2}$. We find that 92.0\% of the CARLA fields are denser than the SpUDS peak density, 55.3\% are $\geqslant$ 2 $\sigma$ overdense and 37.0\% are $\geqslant$ 3 $\sigma$ overdense. At the 5 $\sigma$ level, 9.6\% of the CARLA fields are still overdense. 
The corresponding values for the SpUDS field are significantly lower, with 18.7\%, 8.6\%, and 0.7\% of the fields being denser than $\Sigma_{\rm{SpUDS}}+2\sigma_{\rm{SpUDS}}$,  $\Sigma_{\rm{SpUDS}}+3\sigma_{\rm{SpUDS}}$,  $\Sigma_{\rm{SpUDS}}+5\sigma_{\rm{SpUDS}}$, respectively (see Fig. \ref{histo}). The densest CARLA field, a HzRG at $z = 2.00$, shows a density of almost $\Sigma_{\rm{SpUDS}}+10\sigma$. A two-sided Kolmogorov-Smirnov (K-S) test gives a $ < 10^{-36}$ probability that the CARLA and SpUDS surface density distributions are drawn from the same underlying distribution. In addition, a Mann-Whitney U test rules out that the two distributions have the same mean. 

As the CARLA fields are distributed over the whole sky, we compute the Spearman rank correlation coefficient $\rho$ to test the dependence of density on Galactic latitude to rule out any bias due to stellar contamination. We find $\rho = 0.01$, implying a 21\% probability of a correlation. According to standard interpretations of the Spearman rank correlation test, this is considered as evidence for the lack of a correlation.

\citet{Galametz_2012} also used a counts-in-cell analysis to identify overdensities of IRAC-selected sources among 72 HzRGs in a redshift range of $1.2 < z < 3$, albeit with observations two magnitudes shallower than CARLA. They found that $73 \pm 12 $\% of the fields were denser than the mean blank-field density, derived from the SWIRE survey. They also found that 23 $\pm 7$ \% of the fields were $\geqslant$ 2 $\sigma$ overdense. We cut the CARLA survey at the depth of the \citet{Galametz_2012} study to check for consistency with this previous work. The SpUDS mean blank-field density, derived from a Gaussian fit to the low density half of the distribution, gives $2.1\pm0.7$ IRAC-selected sources per arcmin$^{2}$, in good agreement with the blank field mean density derived by \citet{Galametz_2012}, $\Sigma_{\rm{SWIRE}} = 2.8 \pm 1.0$ arcmin$^{-2}$. Most of the CARLA fields are denser than the blank field, with $79 \pm 9\%$ denser than the SWIRE surface density, consistent with \citet{Galametz_2012}. At the high end of the density distribution we find more fields, with $51 \pm 7 \%$ being $\geqslant$ 2 $\sigma$ overdense. This may, however, be due to small number statistics in \citet{Galametz_2012}
 
In Fig. \ref{histo} we show the position of a known cluster, CIG 0218.3-0510 at $z = 1.62$ \citep{Papovich_2010}, in the density distribution. This cluster lies in the SpUDS survey and is centered on one of our random apertures placed onto the survey. The aperture covering this cluster is the densest one in the SpUDS survey. In \citet{Galametz_2012}, the IRAC color criterion has been successfully tested by recovering known high-redshift (proto-) clusters such as PKS1138$-$262 ($z = 2.16$, the Spiderweb galaxy). In addition, we recently confirmed a cluster in the field of MRC0156-252, a radio galaxy at $z = 2.02$, which is 2 $\sigma$ overdense in the CARLA data (Galametz et al., in prep.). 

These results confirm that the color criterion is very effective at identifying high-redshift clusters and that the densities found in this work are sufficient to qualify many of the fields as promising (proto-)cluster candidates. 
To demonstrate the quality of our data, Fig. \ref{data_img} shows the IRAC2 images of two CARLA sources, the quasar J105231.82+08060 ($z = 1.64$; 6.8 $\sigma$ overdense) and the radio galaxy MRC1217-276 ($z = 1.90$; 5.3 $\sigma$ overdense).  IRAC-selected sources within 1\arcmin\ of the radio-loud AGN are indicated, illustrating the diversity of environments.  While the former shows a central concentration of IRAC-selected sources close to the radio-loud AGN, the latter shows a more diffuse overdensity, perhaps suggestive of an earlier stage in the cluster formation.

\subsection{Radial Distribution of IRAC-Selected Sources}

\begin{figure}
\centering
\includegraphics[trim = 1.1cm 0cm 0cm 0cm, clip = true, scale = 0.45]{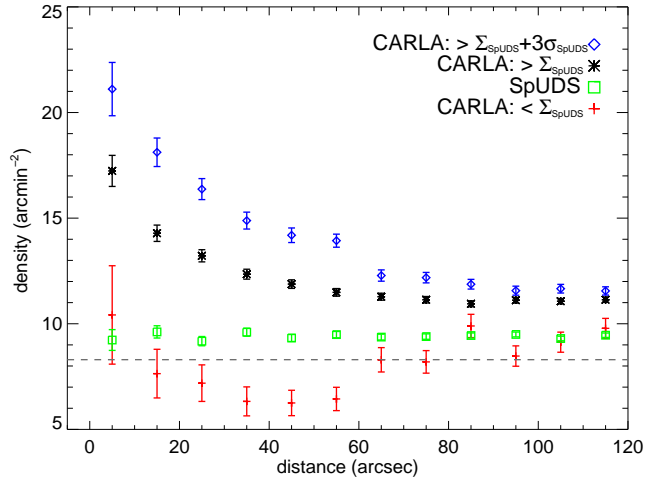}
\caption{Surface density of the IRAC-selected sources as a function of distance to the radio-loud AGN for the CARLA survey. We show a stack of all CARLA fields that are denser and less dense than $\Sigma_{\rm{SpUDS}}$ (black asterixes and red crosses, respectively) and a stack of fields denser than $\Sigma_{\rm{SpUDS}}+3\sigma_{\rm{SpUDS}}$ (blue diamonds). The radial distribution depends on the overall density of the CARLA fields with a sharp rise towards the center for the densest CARLA fields and an almost homogenous distribution for the lowest density fields. The grey dashed line corresponds to the SpUDS peak density and is slightly lower than the SpUDS radial distribution which is computed from a stack of all SpUDS fields, including both the low- and the high-density fields.}
\label{radial_dens}
\end{figure}

We investigate how the surface density of the IRAC-selected sources
depends on the distance to the radio-loud AGN. In Fig. \ref{radial_dens}
we show a stack of the surface density of IRAC-selected sources in
annuli around the radio-loud AGN for both the CARLA and the SpUDS
survey. As expected, the radial distribution of IRAC-selected sources
in the SpUDS survey is homogenous and is equivalent to a median blank
field surface density. It should be noted that this median is
dependent on large-scale structure and existing clusters in SpUDS
and is therefore slightly higher than the typical blank field density
derived from the Gaussian fitting to the lower part of the density
distribution (see Fig. \ref{histo}). For the dense CARLA fields, the surface density peaks
towards the position of the radio-loud AGN in the center of the
image. This rise is even more prominent for the CARLA fields overdense
at the 3 $\sigma$ ($\Sigma = \Sigma_{\rm{SpUDS}}+3\sigma_{\rm{SpUDS}}$)
level than for the CARLA fields that are simply denser than the
peak SpUDS density.
This means that, on average, the denser a field
the more IRAC-selected sources are found very close to the radio-loud
AGN. This result clearly shows that for the dense CARLA fields, the
radio-loud AGN is a good beacon for the center of the overdensity
of IRAC-selected sources.  

We also test that this behavior is not simply due to the
CARLA sample centering apertures on known massive, high-redshift
galaxies while the SpUDS comparison sample targeted random positions
in that field.  We identify a subsample of SpUDS sources with
similar mid-infared colors and magnitude distributions to the HzRGs
in CARLA; the CARLA RLQs are too dominated by their AGN in the mid-infrared to use
for this comparison analysis.  We find that the density of IRAC-selected
sources for this matched sample does not show a prominent peak towards
the center, and is consistent with the SpUDS blank field density
at all radii, thus further supporting our primary result that powerful
radio-loud AGN are efficient beacons for high-redshift structure.

Finally, we also investigate the radial density distribution for all CARLA fields
that are less dense than the SpUDS peak density to rule out 
systematics in our analysis. The rise towards the center fully
disappears and is consistent with the SpUDS peak density shown by
the grey dashed line. It is important to note that even at largest
radii probed, the CARLA fields show an excess of IRAC-selected sources
suggesting that the extent of the overdensity is larger than the
area covered in our images. \citet{Venemans_2007} measures the size
of (proto-)clusters by analyzing the radial extent of Ly$\alpha$
emitters in the fields of radio galaxies at $z \sim 3$.  This work
showed that only at a $\sim$ 1.75 Mpc ($\sim$ 3.8 arcmin) distance from the radio galaxy does the density of Ly$\alpha$ emitters in the field become consistent
with the field density. They conclude that the sizes of (proto-)clusters
as inferred from the distribution of Ly$\alpha$ emitters at $z \sim
3$ is roughly 2 Mpc. Due to the limited size of the CARLA fields
we cannot measure the overall extent of the overdensity but we show
that it extends beyond 2 arcmin ($\sim$ 1 Mpc for the redshifts
probed). The clear rise towards the center at distances $< 60$
arcsec, however, justifies our choice of 1 arcmin apertures for the
counts-in-cell analysis.

\subsection{Dependence on AGN Type}

\begin{figure}
\centering
\includegraphics[scale = 0.5]{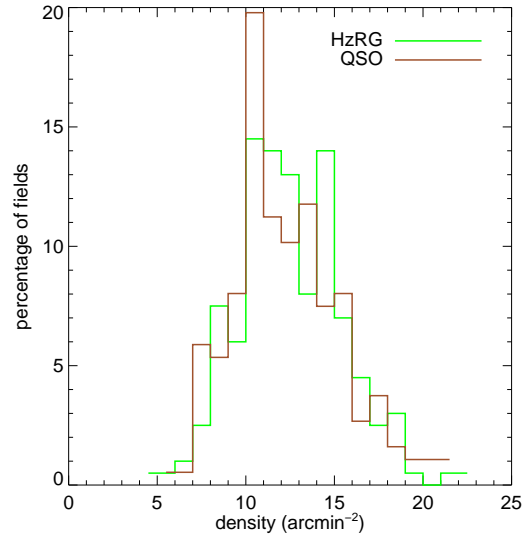}
\caption{Distribution of the densities of IRAC-selected sources for the HzRG and RLQ fields in the CARLA sample. A two sided K-S test shows no significant difference between the two distributions, implying that the environments of the two AGN types are similar.}
\label{histo_rg_rlq}
\end{figure}

We compare the fields around the HzRGs and the RLQs separately to investigate any difference in their environments. Fig. \ref{histo_rg_rlq} shows the density distributions for the HzRGs and the RLQs in our sample. We proceed in the same way as earlier and fit a Gaussian to the low density part of the histograms to find the mean and the standard deviation of the distributions. The best fit is found for a mean surface density of IRAC-selected sources of 11.6$\pm 2.2$ arcmin$^{-2}$ for the HzRGs and 10.8$\pm 1.7$ arcmin$^{-2}$ for the RLQs. The fitted means agree within the uncertainties. A two-sided K-S test shows the two distributions to be statistically similar, with only a 38\% chance of the two distributions being different. The Mann-Whitney U test only gives a 23\% chance of the two means of the distributions being different. We repeat the analysis for various redshift bins to test if this result is only true for specific redshifts, finding similar results for all tested redshift bins. We discuss the implications of environment not correlating with AGN obscuration in Section 5.

\subsection{Dependence on Redshift}

\begin{figure}
\centering
\includegraphics[scale = 0.55]{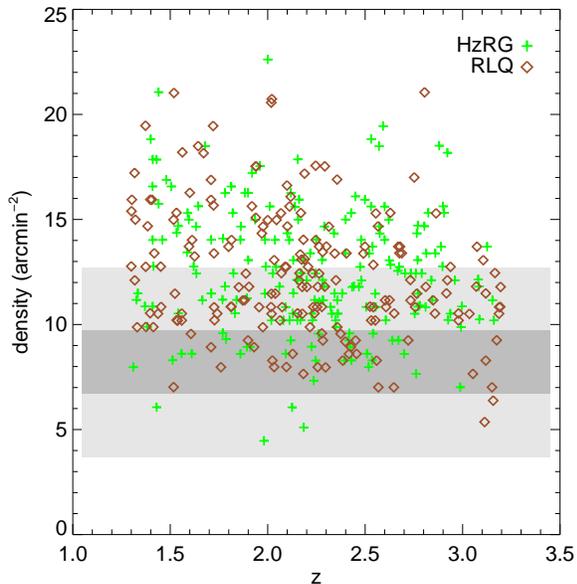}
\caption{Surface density of IRAC-selected sources in the HzRG (green crosses) and RLQ (orange diamonds) fields as a function of redshift. The dark and light grey shaded areas show $\Sigma_{\rm{SpUDS}} \pm1\sigma$ and $\Sigma_{\rm{SpUDS}}\pm3\sigma$ density, respectively. A Spearman rank correlation analysis shows a dependence of surface density with redshift.}
\label{d_z}
\end{figure}

Figure \ref{d_z} shows the environments of radio-loud AGN as a function of redshift. Spearman rank correlation coefficients suggest a dependence on redshift with a probability of 99.99\%. This correlation is mainly driven by fewer data points at $z < 2.0$ and $\Sigma < 10$ arcmin$^{-2}$ and at $z >2.5$ and $\Sigma > 12$ arcmin$^{-2}$. We therefore also test for a correlation in these two redshift bins ($1.3<z<2.5$ and $2.5<z<3.2$); the Spearman rank correlation coefficients suggest a correlation with a probability of 99\% and 82\%, respectively and confirm that the dependence of density with redshift is true for all redshifts probed here. The density of the environments around the RLQs seems to be more dependent on redshift than those of the HzRGs with probabilities of 99.0\% and 83.0\% of a correlation, respectively. However, this could be due to slightly different sampling at intermediate redshifts. If galaxy (proto-)clusters have the same number of members at all redshifts, a dependence with redshift would be expected as we observe galaxies that are about 3 magnitudes fainter than $L^{*}$ at $z \sim 1.3$ but only galaxies that are $\sim$ 2 magnitudes fainter than $L^{*}$ at $z \sim 3.2$. Wylezalek et al. (in prep.) uses the CARLA sample to quantitatively examine the redshift evolution of the cluster luminosity function. The fact that we observe a dependence with redshift demonstrates that a non-negligible fraction of the IRAC-selected objects in the vicinity of the radio-loud AGN are associated with the AGN. If the IRAC-selected sources were not associated with the radio-loud AGN then their density would likely be independent of the AGN redshifts. 

\subsection{Dependence on Radio Luminosity}

\begin{table*}
\scriptsize{
\caption{Comparison to previous work on type-1 and type-2 AGN environments.}
\begin{center}
\begin{tabular}{l c c c c}
\hline\hline
Reference & Sample Size &$z$ & $L_{500\rm{MHz, \ min}}$ & type-1 vs. type-2 environment \\
 & & & (W Hz$^{-1}$) & \\
\hline
\citet{Yates_1989} &  25 RGs &$0.15 < z < 0.82$ & $1 \cdot 10^{25}$ & No difference \\
\citet{Smith_1990} & 35 RGs, 31 RLQs & $< 0.3$ & $5 \cdot 10^{24}$ & No difference \\
\citet{Falder_2010} & 75 RLQs, 27 RGs & $ z \sim 1$ & $ 2 \cdot 10^{24}$ & Slightly denser type-2 environments  \\ 
\citet{Donoso_2010} & 307 RLQ, $\sim$ 1400 RGs& $ 0.4 < z <0.8 $ & $ 2 \cdot 10^{25}$ & type-2 more clustered for  $10^{25} <  L_{500\rm{MHz}} <  10^{26}$,\\
 & & & &  similar clustering for $L_{500\rm{MHz}} > 10^{26} $ \\
This work & 166 RGs, 180 RLQs &$1.3 < z < 3.2$ & $3 \cdot 10^{27}$ & No difference\\
\hline
\end{tabular}
\end{center}
\label{hzrg_rlq}
}
\end{table*}
\begin{figure}
\centering
\includegraphics[scale = 0.55]{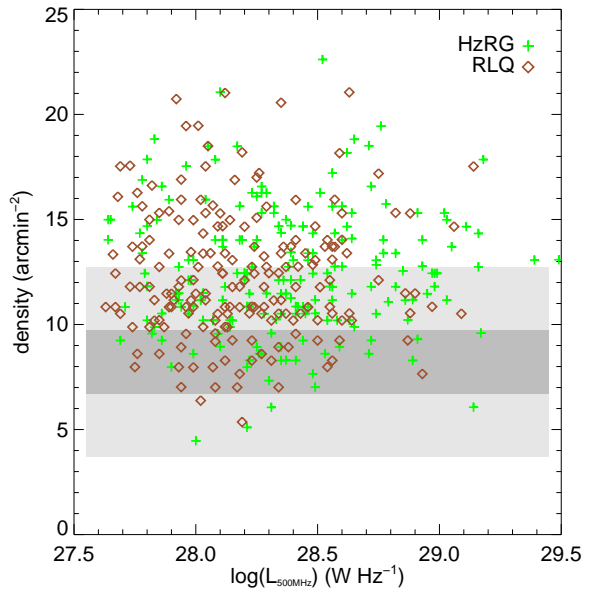}
\caption{Surface density of IRAC-selected sources in the HzRG and RLQ fields as a function of radio luminosity. The dark and light grey shaded areas show  $\Sigma_{\rm{SpUDS}}\pm1\sigma$ and $\Sigma_{\rm{SpUDS}}\pm3\sigma$ density, respectively. A Spearman rank correlation analysis shows no significant dependence of surface density with radio luminosity.}
\label{d_radio}
\end{figure}
Figure \ref{d_radio} shows how the density of the CARLA fields depends on the radio luminosity of the targeted radio-loud AGN. Spearman rank correlation coefficients do not show a significant dependence of the CARLA field densities on $L_{500\rm{MHz}}$. This is in agreement with previous studies at these high radio luminosities \citep{Donoso_2010} and is consistent with expectations from the unified model for AGN \citep{Urry_1995}. We discuss this result in the context of other studies at lower redshifts in Section 6.

We also test the difference of the density distributions between the HzRG and RLQ fields as a function of radio luminosity. We fit a Gaussian to the density distribution in four radio luminosity bins ($10^{27}-10^{28}$ W Hz$^{-1}, 10^{28}-10^{28.25}$ W Hz$^{-1}, 10^{28.25}-10^{28.5}$ W Hz$^{-1}, 10^{28.5} -10^{29}$ W Hz$^{-1}$, chosen to provide similarly sized samples) to determine the mean and sigma of the histograms. Two-sided K-S tests do not show a significant probability for the distributions to be different. Spearman rank correlation coefficients do not imply a dependence of the density on radio luminosity for either of the two types of AGN.  

\section{The environments of HzRGs and RLQs}

It has long been debated whether the environments of radio-quiet AGN are any different from their radio-loud analogs. Many studies show that radio-quiet AGN, which are usually associated with less massive spiral galaxies, are less clustered than radio-loud AGN \citep[e.g.,][]{Hutchings_1999, Kauffmann_2008, Hickox_2009, Falder_2010}. Several theories have been proposed to explain this difference between the environments of radio-loud and radio-quiet AGN.  One theory, known as jet confinement, posits that the higher IGM densities in regions of higher galaxy densities enhance synchrotron losses from radio jets, thereby making them brighter \citep{Barthel_1996}. An alternative theory suggests that mergers in regions of higher galaxy density will produce more rapidly spinning supermassive black holes, which will then be more capable of powering a radio jet \citep[e.g.,][]{Wilson_1995, Sikora_2007}. Regardless, observational results suggest that radio-quiet AGN reside in less dense environments while the difference between RLQ and RG environments still remains an open issue and observations show inconclusive results. 

From a study of $\sim$ 30 radio-selected AGN at $z<0.3$, \citet{Smith_1990} finds that RLQs and RGs at $z < 0.3$ were similarly clustered, consistent with the results of \citet{Yates_1989} based on a comparably sized sample at similar redshift (see Tab. \ref{hzrg_rlq}). \citet{Falder_2010} studies the environments of 75 RLQs and 27 RGs at $z \sim 1$, and finds a small, but notable trend of density with AGN type, with the RG environments being slightly denser than the RLQ environments. He suggests, however, that this might be due to the different sample sizes. \citet{Donoso_2010} computes the cross-correlation signal between a sample of $\sim$ 14,000 radio-detected AGN and a sample of 1.2 million luminous red galaxies in the SDSS. They also study a sample of $\sim$ 300 RLQs cross-correlated with the same sample of luminous red galaxies. When comparing the two clustering signals, \citet{Donoso_2010} finds the RGs are more clustered than the RLQs for $10^{25} < L_{1.4\rm{GHz}} < 10^{26}$ W Hz$^{-1}$, but clustered in a similar way for $L_{1.4\rm{GHz}} > 10^{26}$ W Hz$^{-1}$. 

We caution the reader that in each study different lower limits and definitions of radio-loudness are being used. In Tab. \ref{hzrg_rlq} we summarize the results of these works and list the redshift ranges and the lower limit on radio luminosity used to define a radio-loud object. The radio luminosities have been converted to $L_{500\rm{MHz}}$ assuming a radio spectral index of $\alpha = -0.7$  \citep{Miley_2008}. At lower radio luminosity, $L_{500\rm{MHz}} < 10^{25}$ W Hz$^{-1}$, the studies show somewhat conflicting results. A concerning problem is the small sample sizes providing poor statistics and with imperfectly matched samples. In \citet{Falder_2010}, for example, the RGs were slightly more radio luminous than the RLQs.

At low radio luminosities, our understanding of the environments of RLQs and RGs is not yet clear. At higher radio luminosities, $L_{500\rm{MHz}} > 10^{25}$ W Hz$^{-1}$, previous studies and our analysis agree that there are no significant differences between the environments of type-1 and type-2 AGN.  These results are consistent with the popular orientation-driven (e.g., obscuring torus) AGN unification scenario for the most powerful type-1 and type-2 radio-loud AGN.  However, at lower radio luminosities, other factors, such as AGN feeding mechanisms, appear to come into play causing less powerful radio-loud type-1 and type-2 AGN to, on average, reside in different environments.

\section{Summary}

We have studied the environments of 387 radio-loud quasars and radio galaxies at $1.3 < z < 3.2$ that have been observed at $3.6 \mu$m and $4.5\mu$m with the \textit{Spitzer Space Telescope}. Our study clearly confirms that radio-loud AGN reside in rich environments. By applying a color cut, we identified likely high-redshift galaxies, which we refer to as `IRAC-selected sources', and computed their surface density in a $1^{\prime}$ radius around the radio-loud AGN. More than half of the fields, 55.3\%, are overdense at the 2 $\sigma$ level when compared to the blank field density of IRAC-selected sources derived from the SpUDS survey. If we instead consider a 5 $\sigma$ overdensity, 9.6\% of the fields are overdense. We detect a significant rise in surface density of IRAC-selected sources towards the position of the radio-loud AGN in contrast to a homogeneous radial surface density distribution for the SpUDS apertures. This reassures us of an association of the majority of the IRAC-selected sources with the radio-loud AGN. These CARLA results provide a promising list of high-redshift galaxy (proto-)cluster candidates for further study. 

We also study the dependence of environment on radio luminosity. No statistically significant correlation is observed. We do, however, find a significant dependence on redshift, with higher redshift fields being slightly less dense. This result is consistent with galaxy cluster evolution and reassures us that most of the IRAC-selected sources are indeed associated with the targeted radio-loud AGN as otherwise we would have been unlikely to have observed a redshift dependence. In Wylezalek et al. (in prep.) we discuss how this trend can be explained in the context of the evolution of the luminosity/mass function of galaxy clusters.

We do not find a significant difference between the RLQ and HzRG fields, neither in their overall distribution nor in their dependence with redshift or radio luminosity. We conclude that for high radio luminosities, $L_{500\rm{MHz}}> 10^{27.5}$ W Hz$^{-1}$, orientation-driven unification models appear valid. 

\acknowledgments

We gratefully acknowledge James Falder and Conor Mancone who were involved in the initial CARLA proposals. NS is the recipient of an ARC Future Fellowship. This work is based on observations made with the \textit{Spitzer Space Telescope}, which is operated by the Jet Propulsion Laboratory, California Institute of Technology under a contract with NASA.

\clearpage

\end{document}